\documentclass[a4paper,11pt]{article}
\usepackage{pos}
\usepackage{graphicx}
\usepackage{makecell}

\title{GPU performance in Run3 ALICE online/offline reconstruction}

\author*[a]{Gabriele Cimador}
\onbehalf{on behalf of the ALICE collaboration}

\affiliation[a]{Dipartimento di Matematica, Informatica e Geoscienze, Università degli studi di Trieste and\\INFN sezione di Trieste,\\34127 Trieste, Italy}

\emailAdd{gabriele.cimador@cern.ch}

\abstract{After several software and hardware upgrades during LS2, ALICE records 50 KHz of minimum bias Pb--Pb collisions in continuous readout mode. To cope with the high data rate of 3.5 TB/s from the detectors, multiple stages of compression are employed during data taking, the last one requiring full TPC tracking. This compression chain is part of the synchronous (online) processing and largely relies on GPU computing to reduce the data rate to 170 GB/s.
After the data taking, a second phase called asynchronous (offline) processing is performed, where the compressed data are read and a full global reconstruction is performed, which results in the production of data that can be analysed. When the online computing farm is not fully utilized for online processing, it is capable of running the offline reconstruction leveraging GPUs, which proved to speed up also this phase. ALICE is making significant efforts to offload more of the asynchronous reconstruction to GPUs, to speed up the overall execution time and efficiently use as much GPU computing resources as possible. This talk will focus on the performance that GPUs have provided in Run 3 for the ALICE experiment, both for online and offline reconstruction. Moreover, it will also highlight the offloading process to GPU of the track-model decoding, an asynchronous reconstruction task which is one of the software routines that have been targeted to be offloaded to GPUs during offline reconstruction.}

\FullConference{12th Large Hadron Collider Physics Conference  (LHCP2024)\\
3-7 June 2024 \\
Boston, USA \\}

\begin{document}
\maketitle

\section{Introduction}
During Long Shutdown 2, ALICE \cite{alice-experiment} underwent several software and hardware upgrades. The upgrades allowed ALICE to operate in continuous readout recording up to 50 KHz of minimum bias Pb--Pb collisions, 100 times more than Run 2. The increased collision rate and continuous readout lead to a significant increase of the data rate coming from the detectors, reaching 3.5 TB/s during data taking. To avoid to store all this data, a chain of compression stages executed by two computing farms reduces the data rate on the fly, cutting down the rate to $170$ GB/s. After a low
level compression via zero-suppression done by the first farm, the Event Processing Nodes (EPN) farm performs advanced data reduction to further decrease the data rate and extract calibration data.\\
A lossless Asymmetric Numeral Systems (ANS) entropy encoding \cite{lettrich-entropy} is used to compress the data of all detectors. Since the Time Projection Chamber (TPC) is responsible for more than 90\% of the total data size, several processing steps are employed before the ANS compression to reduce the TPC data size \cite{rohr-global-track}.
The result of this compression chain is the calibration data and the Compressed Time Frames (CTFs) which contain collisions data. These objects are stored into a temporary disk buffer and later are transferred to the Tier 0 and Tier 1 of the Worldwide LHC Computing Grid (WLCG). This data processing phase is called Online or Synchronous because everything happens during data taking.\\
A second data processing phase called Offline or Asynchronous is later executed, where the CTFs are decompressed and the global tracks are reconstructed using the calibration data extracted during the online phase. The final result of this phase is a set of Analysis Object Data files (AO2D), the actual data on which the ALICE community performs the physics analysis.\\
Alongside the hardware, software has also been upgraded during LS2, developing a new computing system named O$^2$, standing for Online-Offine (framework). The framework combines detector read-out, event building, data recording, detector calibration, data reconstruction (both online and offine), physics simulation and analysis; it integrates the online and offine processing under the same codebase, meaning that the algorithms are employed for both the online and the offline reconstructions. O$^2$ was engineered to run on heterogeneous systems, and part of the routines of O$^2$ must be able to run both on CPUs and hardware accelerators seamlessly. The reconstruction software stack of O$^2$ thus includes a framework specialized in General-Purpose GPU computing (GPGPU), which is compatible also with CPU execution. The same source code, written in generic \texttt{C++} is compatible to CUDA, OpenCL, HIP and CPU parallel computing through OpenMP \cite{rohr-lindenstruth-hpccws}.\\
The online processing executed by the EPN must keep up with the 900 GB/s rate sent by the first farm. Since Run 2, when GPUs were used in the High-Level Trigger (HLT) farm, they have proven to be optimal for fast track reconstruction, becoming essential for the ALICE experiment from the very beginning. As a result, the majority of the computing power in the EPNs was chosen to be in the form of GPUs, with a significant increase of their importance for Run 3 and Run 4 compared to the HLT farm used in Run 2. To ensure high computing throughput, each EPN is equipped with 8 high-end AMD GPUs, for a total of 2800 GPUs (350 nodes). Offline processing, on the other hand, can leverage the GPU’s computing capacity when the EPN is not fully occupied with online processing. Thus, GPUs play a fundamental role in the overall ALICE data processing plan.

\section{GPU performance during online processing}
\begin{figure}[h]
    \centering
    \includegraphics[width=0.75\textwidth]{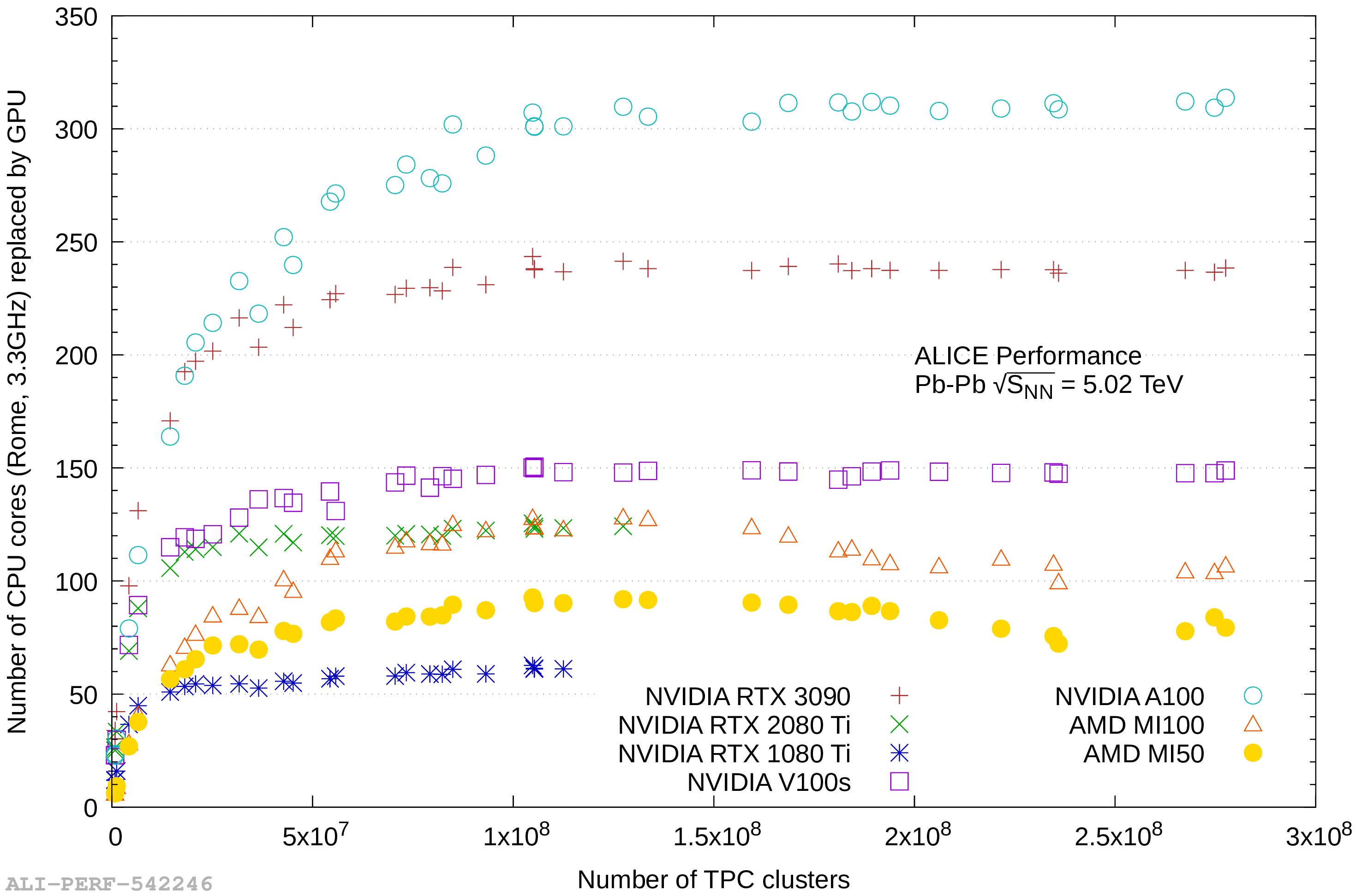}
    \caption{Speedup of the O2 GPU synchronous reconstruction compared to a single CPU core.}
    \label{fig:online_speedup}
\end{figure}

The EPN thus need to sustain a input data rate of 900 GB/s coming from the first computing farm. The TPC account for 90\% of total data, requiring ad-hoc processing steps to further reduce its data size. These processing steps necessitate the full TPC track reconstruction for all the collisions inside the TPC volume. As a result, the TPC processing takes 99\% of synchronous time as shown in table \ref{tab:sync}. The TPC processing comprises the clustering of raw data, full TPC reconstruction and compression. In addition, the online processing needs full barrel reconstruction for a small fraction of the events to extract calibration data. Consequently, the EPN farm has been specifically tailored to run the fastest TPC reconstruction possible on GPUs in order to keep up with the 900 GB/s input data rate. Using Monte-Carlo simulation data as input of the reconstruction workflow, it was possible to study how many CPU cores a GPU can replace in TPC online processing, the most critical task of the online phase. Figure \ref{fig:online_speedup} shows the speedup of the online TPC processing normalized to a single EPN CPU core. Simulated data shows that GPUs can replace 50-300 EPN CPU cores for ALICE TPC online processing. The EPN was initially configured with 250 nodes with 8 AMD MI50 GPUs each, since those GPUs provided the best cost vs performance ratio at the time. After the tests on 2022 Pb--Pb data, the EPN farm was extended, as the online processing required more computing resources than expected to cope with the 50 kHz interaction rate of Pb--Pb collisions. This expansion involved the addition of 30 more nodes with the same GPU configuration, along with 70 additional nodes, each equipped with 8 AMD MI100 GPUs.
It is worth noting that with CPU computing capacity only, the EPN farm would have required around 3000 nodes with related networking and maintenance. On the other hand, by using GPUs, the computing farm needed 350 nodes to achieve the required computing power for the online processing, including 30\% safety margin of compute workload. As a result, GPUs proved to be the most efficient, cost-effective and easier to maintain solution, helping to keep the project within budget.
\section{GPU performance during offline processing}
The offline processing regards the full reconstruction with calibration, including the reconstruction of global tracks across the detectors, identification of the primary and secondary vertices, and the assignment of the particle identification (PID) hypotheses. The compute load is split on the GRID and on the EPN when the LHC is not in operation or during pp data taking at 650 kHz in which not all resources are needed for online processing. Despite the online and offline phases share some common routines, the relative time taken by each detector processing steps are different. Table \ref{tab:sync} shows the percentage of Linux CPU time taken by the online processing steps, while table \ref{tab:async} highlights offline processing steps. It is important to note that table \ref{tab:sync} shows only reconstruction tasks and leaves out calibration, quality control, event building and  network IO. However, TPC processing still consistently accounts for over 90\% of the online processing when considering the total CPU workload. The TPC processing takes a smaller portion in the offline processing, since the clustering and compression are not required and the full reconstruction also for other detectors is performed. When the EPN executes the offline processing, 60\% of it is executed by the GPUs, i.e. the TPC processing. This offloading leads to a speedup by a factor of 2.5 \cite{eulisse-rohr-o2}. ALICE computing plans foresee to offload the full central barrel global tracking chain to GPUs; this routines include Inner Tracking System (ITS), Time Projection Chamber (TPC), Transition Radiation Detector (TRD), and Time-Of-Flight (TOF) detector tracking and matching, secondary vertexing and track refitting. Given that the full tracking takes around 80\% of the total offline time, this offload should lead to a speedup by approximately a factor of 5 when executing the offline reconstruction on the EPN farm.
\begin{table}
    \centering
    \begin{minipage}{0.45\textwidth}
        \centering
        \caption{Relative processing times of synchronous reconstruction steps. 50 kHz Pb--Pb Monte-Carlo
simulated data \cite{eulisse-rohr-o2}.}
        \begin{tabular}{l r}
    \hline
    Processing step & Relative time \\
    \hline
    \makecell[l]{TPC Processing (Clustering, \\Tracking, Compression)} & 99.37\% \\
    EMCAL Processing & 0.20\% \\
    \makecell[l]{ITS Processing \\ (Clustering + Tracking)} & 0.10\% \\
    TPC rANS Encoding & 0.10\% \\
    ITS--TPC matching & 0.09\% \\
    MFT Processing & 0.02\% \\
    \makecell[l]{TOF Processing and \\ Global Matching} & 0.02\% \\
    \hline
    Rest & 0.1\% \\
    \hline
\end{tabular}
        \label{tab:sync}
    \end{minipage}
    \hfill
    \begin{minipage}{0.45\textwidth}
        \centering
        \caption{Relative processing times of asynchronous
reconstruction steps. 650 kHz pp real data, 2022, no
calorimeters in run \cite{eulisse-rohr-o2}.}
        \begin{tabular}{l r}
            \hline
            Processing step & Relative time \\
            \hline
            TPC Processing (Tracking) & 61.41\% \\
            ITS-TPC matching & 6.36\% \\
            MCH & 6.13\% \\
            TPC Entropy Decoding & 4.65\% \\
            ITS Tracking & 4.16\% \\
            TOF Matching & 4.12\% \\
            TRD Tracking & 3.95\% \\
            MCH Tracking & 2.02\% \\
            AOD Production & 0.88\% \\
            \hline
            Quality Control & 4.00\% \\
            \hline
            Rest & 2.32\% \\
            \hline
        \end{tabular}
        \label{tab:async}
    \end{minipage}
\end{table}
\section{TPC track-model decoding on GPU}
\begin{figure}
    \centering
    \includegraphics[width=0.85\textwidth]{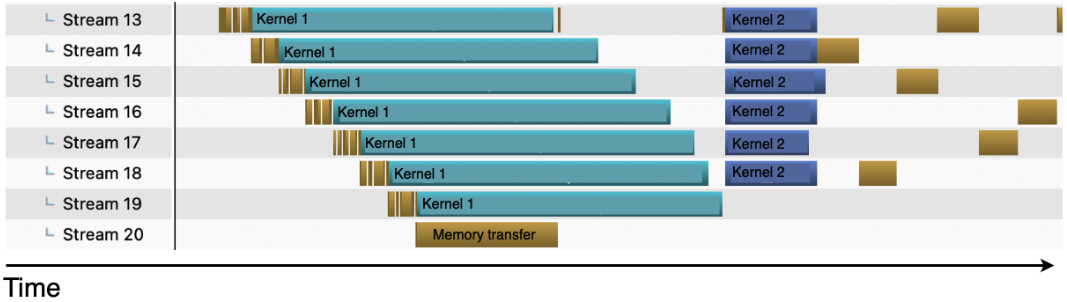}
    \caption{Timeline of the GPU decoding. In yellow memory transfer between CPU and GPU, light blue is the attached clusters kernel, blue is for the unattached clusters. Every row is a different GPU stream.}
    \label{fig:timeline}
\end{figure}

\begin{figure}[htbp]
    \centering
    \begin{minipage}{0.49\textwidth}
        \centering
        \includegraphics[width=\textwidth]{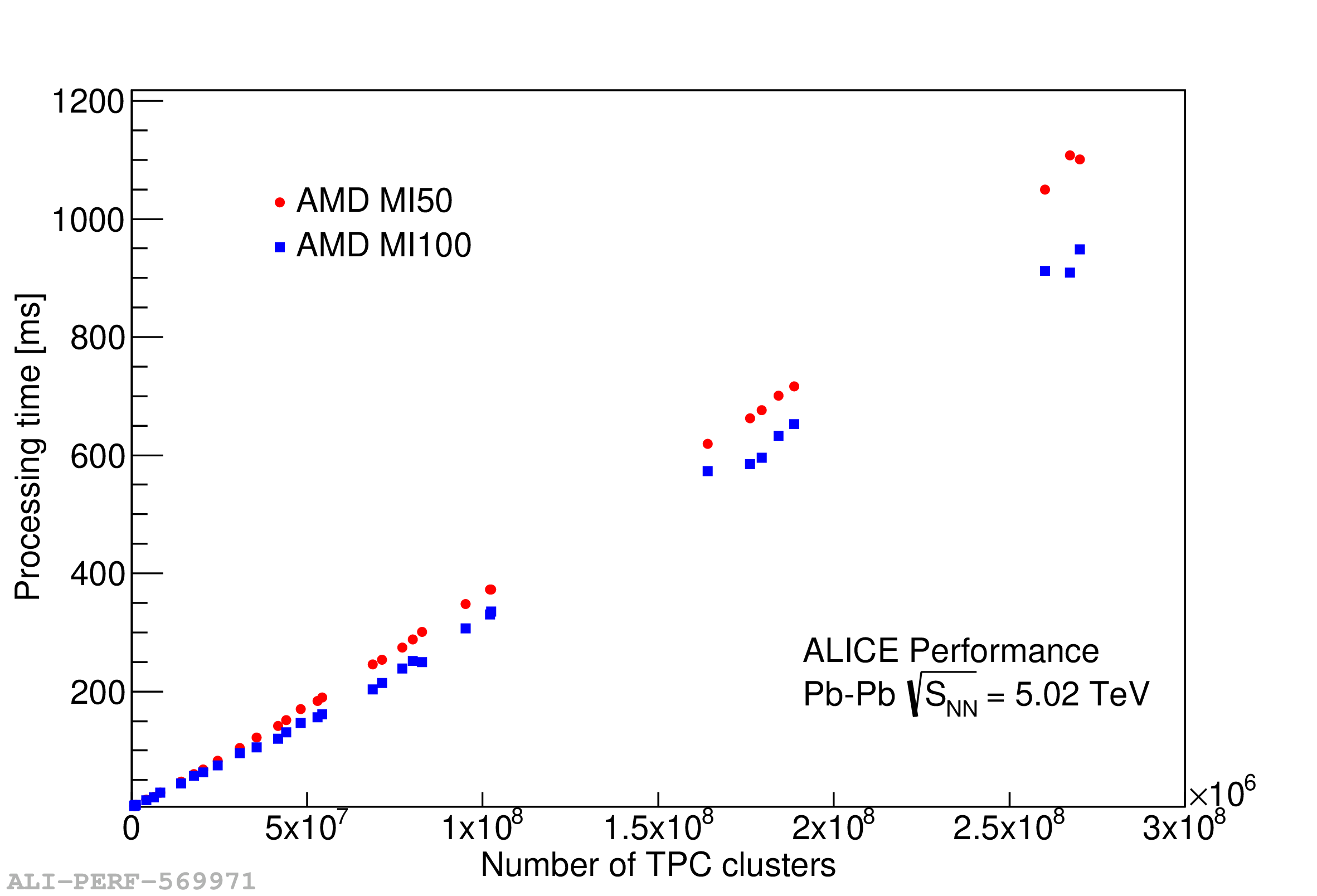}
        \caption{TPC track-model decoding execution times on EPN GPUs.}
        \label{fig:asympt}
    \end{minipage}\hfill
    \begin{minipage}{0.49\textwidth}
        \centering
        \includegraphics[width=\textwidth]{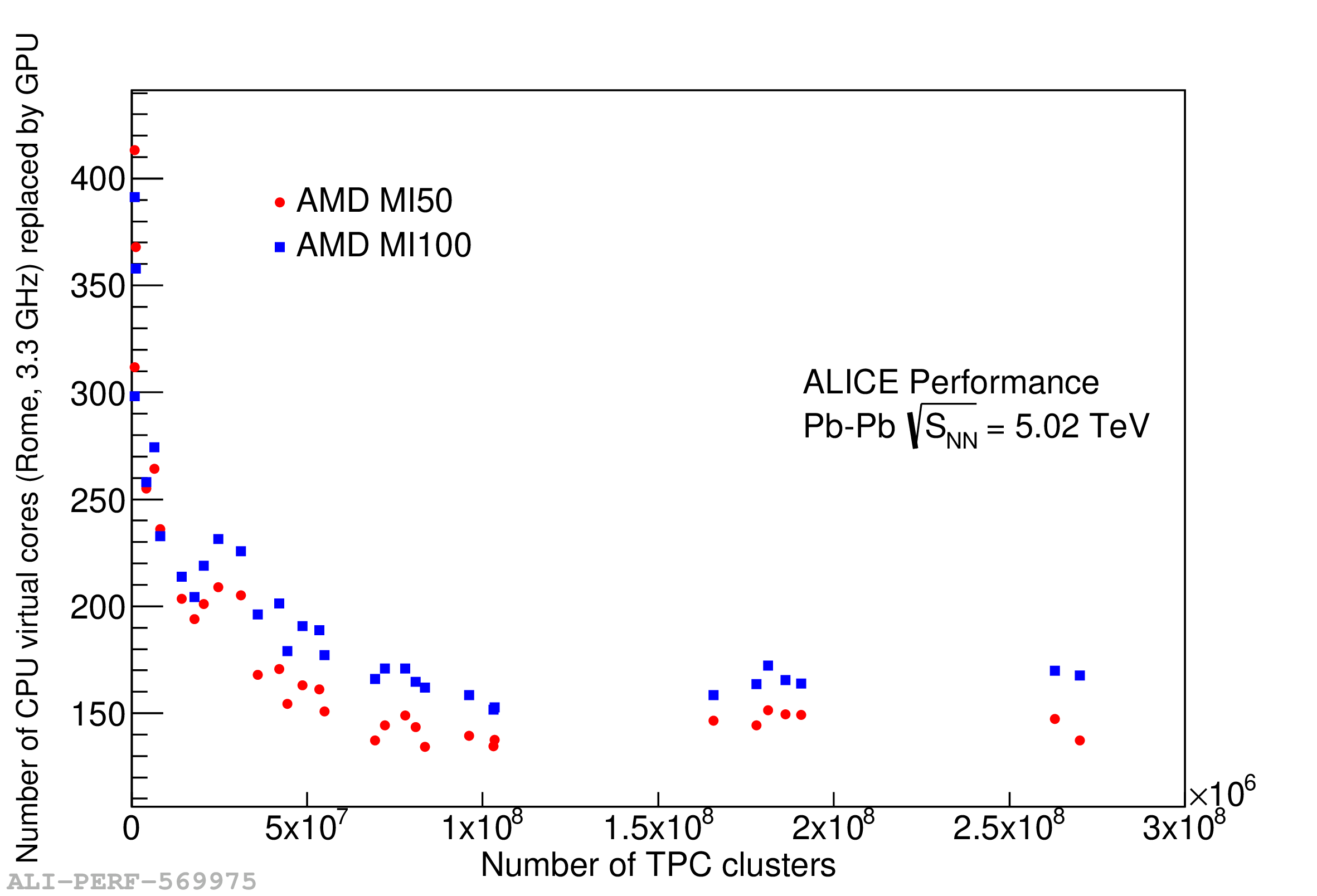}
        \caption{TPC track-model decoding speedup on EPN GPUs, normalised to a single EPN virtual core.}
        \label{fig:speedup}
    \end{minipage}
\end{figure}

All detectors' data are compressed via an entropy encoder \cite{lettrich-entropy}, and thus reducing the entropy of the data would yield a better compression factor. For this reason, before the compression, several preprocessing steps are employed to reduce the entropy of TPC data. One of these steps is the \textit{track-model encoding} \cite{rohr-global-track}. This algorithm aims at encoding the geometrical coordinates of the clusters inside the TPC volume. At the time of the online compression, the full TPC tracking information for all the collisions within the TPC volume is available. Thus, instead of storing the absolute value of the position of the cluster, this algorithm stores each coordinate as the residual to the correspondent track. After the online tracking, some clusters might remain unassigned to any tracks ("unassigned clusters"). To deal with that, the unassigned clusters are sorted by coordinate position inside the TPC volume and their coordinates are stored as differences relative to the preceding cluster. When the offline processing runs, after the entropy decoding, the clusters' coordinates must be decoded to the absolute values instead of residuals or differences, thus the \textit{track model decoding} is executed to restore the original values. This algorithm was originally written for CPU processing and has now also been ported to GPU. \\
The new algorithm consists of two GPU kernels. In the first one, every GPU-thread receives one track, it propagates it through the TPC and computes the absolute values of the coordinates using the residuals given in input and the reconstructed track. The second kernel, for the unattached clusters, computes each coordinate summing the difference taken from the input array with the coordinates of the last decoded cluster; each thread decodes clusters from a specific region of the TPC. Figure \ref{fig:timeline} shows a timeline of the new GPU track model decoding algorithm. 
For each kernel, multiple instances are executed concurrently on different streams, reducing the memory transfer latency. Moreover, some indices to access the input Structure Of Arrays (SOA) of the algorithm have been precomputed on the CPU allowing for a reduction in thread divergence within the GPU routine.\\
When running on the EPN nodes, a single GPU decoding showed no strong superlinear effects when scaling the number of clusters to decode, as shown in Figure \ref{fig:asympt}. Moreover, Figure \ref{fig:speedup} shows the number of virtual EPN cores needed to replace a single EPN GPU for the track model decoding; a single AMD MI50 (MI100) can replace around 140 (170) EPN virtual cores.\\
The porting of the track model decoding on GPUs shows an improvement in the whole offline reconstruction chain. The mean time to process a single timeframe by the whole chain showed to be $2.8\%$ faster for 2023 real pp data and $1.2\%$ faster for real 2023 Pb--Pb data.
\section{Conclusions}
ALICE heavily relies on GPU computing for Run3. The EPN computing farm is equipped with 2800 high-end GPUs which provide the necessary computing capacity to enable the compression and the storage of all heavy-ion collision data in continuous readout. When the EPN is not fully utilised for data readout, the asynchronous processing is executed on this computing farm. Around 60\% of the offline reconstruction is executed on the GPUs, allowing to obtain a speedup by a factor of 2.5 when utilising the EPN. ALICE plans to port the global central barrel reconstruction to GPU computing, which constitutes around 80\% of the offline reconstruction, aiming at achieving a speedup by a factor of 5. Not only this would lead to a better utilisation of the computing power of the EPN, but also it could leverage GPU resources on the GRID if available. Currently efforts are being made to port such additional routines on GPU computing. The track model decoding algorithm is part of these routines and have been successfully ported to GPUs. This porting resulted in an overall improvement of the offline reconstruction chain by 1.2\% for Pb--Pb data and 2.8\% for pp data in terms of timeframe processing latency.


\begin{thebibliography}{99}
\bibitem{alice-experiment}
ALICE Collaboration,
\emph{The ALICE experiment at the CERN LHC}, 
J. Inst. 3, S08002 (2008).

\bibitem{lettrich-entropy}
M. Lettrich,
\emph{Fast Entropy Coding for ALICE Run 3}, 
Feb. 2021, p. 913. 
doi: \url{10.22323/1.390.0913}.

\bibitem{rohr-global-track}
D. Rohr for the ALICE Collaboration,
\emph{Global Track Reconstruction and Data Compression Strategy in ALICE for LHC Run 3}, 
Proceedings of CTD2019 (2019), 
arXiv:1910.12214.

\bibitem{eulisse-rohr-o2}
G. Eulisse and D. Rohr, 
\emph{The O2 software framework and GPU usage in ALICE online and offline reconstruction in Run 3}, 
arXiv:2402.01205 [physics.ins-det] (2024).

\bibitem{rohr-lindenstruth-hpccws}
D. Rohr and V. Lindenstruth, 
\emph{Portable and Vendor-Independent Low-Level Programming and Performance Benchmarking for Graphics Cards and Processors}, 
Proceedings of the 2017 IEEE 19th HPCCWS (2017).

\end{thebibliography}
\end{document}